# Hydrogen bonds and dynamics of liquid water and alcohols

Alexander Kholmanskiy

Science Center «Bemcom», Moscow, Russia

allexhol@ya.ru, http://orcid.org/0000-0001-8738-0189

**Abstract**

Using modified Arrhenius approximations, the activation energies of water, alcohols, and hexane structure rearrangement reactions responsible for the temperature dependences of their dynamic and dielectric characteristics were determined. The interactions of van der Waals and charged centers of water and alcohol molecules regulate the translational and rotational motion of molecules, ensuring the coordination and balance of the thermal effects of exothermic and endothermic rearrangement reactions in local structure of hydrogen bonding network. The long-range action of fluctuating dipoles of hydrogen bonds and their resonant excitation by thermal energy underlies the anomalies in the temperature dependences of water properties and initiates its phase transitions at points 273 K and 298 K. The deviation of the molecular dynamics of water from the Arrhenius and Stokes-Einstein equations in the range of 273-298 K was associated with a high contribution of the collective dynamics of the ice-like phase of water consisting of a network of hydrogen bonds structured by hexagonal clusters of Ih-ice.

## 1. Introduction

The physical nature of the abnormal properties of liquid water is still far from being completely understood [1-5]. More than 100 years of intensive experimental studies and simulations of the molecular dynamics of water have not still answered the question: *«what are the structure and dynamics of the hydrogen bonding network in water that give rise to its unique properties»* [4]. This network of hydrogen bonds (HBs) lies at the root of dynamic supramolecular structure (SMS) of water. Specifics of SMS physics underlie abnormal properties of liquid water, which provide at normal conditions, the existence of biological systems [6-10]. Seeds of many plants get activated in moist environment in vicinity of temperature (T) 277 K (stratification) and have maximal values of germination rate in range of 293-298 K [9]. At normal pressure and in T range from 273 K to ~320 K, water-containing protein systems retain their molecular structure and dynamic properties [10]. Therefore, special attention was paid to the analysis of changes in the dynamic and dielectric properties of water in the range of 273-300 K. For comparison of analysis results, the boundaries of the temperature dependences (TDs) were sometimes expanded to ~373 K and ~250 K. It was taken into account that the spatiotemporal heterogeneity of supercooled water and amorphous ice served as the basis for the hypothesis of the coexistence of two fluctuating phases with high and low density in ambient water [5, 8, 11]. This to two-state mixture model is still being used to explain the anomalies of ambient water [3, 4, 5, 8, 12]. However, in [13, 14] the account of water dissociation reaction initiated due to oxygen atom ionization by ~535 eV photon made it possible to build a computer model for homogeneous structure of water as an alternative to the two-state mixture model.

Rearrangements of SMS water play a key role in phase transitions and in the molecular physics of anomalies TDs of water properties [5, 6, 15, 16]. Arrhenius approximations (ARA) for these TDs are commonly accepted to be a standard method of studying activation energies ($E_A$) of reactions prevailing in rearrangements of water molecular structure. It was found out in [6, 16-18] that for such dynamic characteristics of water as viscosity ($\eta$), self-diffusion coefficient (D), dielectric-relaxation time ($\tau_D$) values of $E_A$ are comparable with the average energy HB ($E_H$) [5]. Besides, these $E_A$ change abruptly in the vicinity of 298 K, and $E_A$ for D are negative, and for $\eta$ and $\tau_D$ are positive. The $E_A$ signs correspond to the thermal effects of endothermic or exothermic reactions of restructuring of the water structure [18].

Such jumps of $E_A$ values lead to the deviation of TDs for dynamic characteristics from Arrhenius as reported in works [19-24], for temperatures T<300 K. At the same time, for interpolation and approximation of experimental TDs, as a rule, power functions T are used, whose numerical parameters are in no way related to the molecular physics of water. Examples of such interpolations are formulas for $\eta$ and the stationary dielectric constant ($\varepsilon_s$) [19, 20]:

$$\eta = \sum_{i=1}^{4} a_i (\frac{T}{300})^{b^i}, \qquad (1)$$

$$\varepsilon_s \text{ (T °C)} = 87.9144 - 0.404399T + 9.58726 \times 10^{-4} T^2 - 1.32892 \times 10^{-6} T^3. \qquad (2)$$

In (1 $a_i$ and $b^i$ the numerical parameters and the error of both interpolations in the range 0-100 °C are of the order of 1%. The deviations of the Stokes-Einstein (SE) and Stokes-Einstein-Debye (SED) equations from linearity at T <300 K compensate by the fractional equations [21-26]:

$$D/T \propto (1/\eta)^t, \qquad (3)$$

$$\tau \propto (\eta/T)^\zeta. \qquad (4)$$

The scatter of the parameters t and $\zeta$ is determined by the measurement accuracy of D and $\tau$ by different methods and in different T intervals. The error in measuring D reaches 9% [27], and the value of $\tau$ can have different physical meanings, depending on the method of probing relaxation processes. In [26, 28], in (4), $\tau_D$ appears as a characteristic of the relaxation of molecular ensembles, but in [22, 25, 29], $\tau_D$ is equated to the reorientation times ($\tau_r$) of an individual molecule. Useful information on the molecular dynamics (MD) of the HBs network can be obtained by comparing the $E_A$ reactions responsible for the TDs of the dynamic and dielectric properties of water, alcohols, and non-hydrogen bonded liquids.

In the current work, in order to identify the features of the molecular physics of water in the range from 273 K to 298 K, a comparative analysis of the temperature dependences of the structural and dynamic characteristics of water, homologous series of alcohols, and hexane was carried out using physically adequate Arrhenius approximations.

## 2. Method and material

It was found out in work [18] that Arrhenius approximations ($F_A$) of temperature dependences of 15 physical characteristics of water can be expressed by a bimodal function:

$$F_A = \exp(\pm E_A/RT) = T^{\pm\beta} \exp(\pm E_R/RT), \qquad (5)$$

where $T^{\pm\beta}$ c $\beta = 0, \pm\frac{1}{2}, \pm 1$ corresponds to the exponential prefactor and the effective activation energy $E_R$ represents the total thermal effect of the rearrangement of the HBs network responsible for the features of the TDs of water characteristics [18]. The $\beta$ values for dynamic characteristics and $\varepsilon_s$ are selected taking into account the known equations of molecular physics. These include SE and SED equations, the Kirkwood formula $\varepsilon_s \propto 1/T$ [20], the Clapeyron equation $P \propto \rho/T$ (P is the pressure, $\rho$ is the density), as well as the relations $\tau_D \propto \tau_r$ [22, 25]. For $\eta$, D, $\tau_D$ and $\varepsilon_s$ the $\beta$ values are: 0, 1, -1, and -1, respectively [18]. Taking into account that the dependence of the dipole moment of the molecule ($\mu$) on T is the derivative TD of the local electric field [30], for $\mu$ we took $\beta = 0$ and $E_A = E_R$, as for $\eta$.

To separate $E_R$ from $E_A$, the function $T^{\pm\beta}$ in (5) was presented as $\exp(\pm E_T/RT)$ and obtained:

$$\pm E_A = \pm E_R \pm E_T.$$

The $E_T$ value in the range 273-298 K was 2.37 kJ/mol and corresponded to thermal energy at T=285.2 K. Approximations (5) were used to analyze the TDs amplitude of fluctuations in the HB angle ($\delta°$) and oxygen – oxygen radial distribution functions $g_{oo}$ (r) of the first ($^1Cx$) and second ($^2Cx$) configurational shells of the central molecule. To evaluate the contribution of van der Waals (vdW) interactions to TDs in [15], two versions of the tetrahedral water network simulation were used. The density functional BLYP corresponds to too rigid HBs, and RPBE variant – very weak HBs. The $\beta$ values for $\delta°$ and $g_{oo}$ (r) were taken as for D and $\eta$, equal to 1 and 0, respectively.

In calculating $E_A$ and $E_R$ from the known TDs of the dynamic and dielectric properties of alcohols from methanol to 1-decanol and hexane, the $\beta$ values adopted for water were used. When plotting the dependence of $\varepsilon_s$ for liquids with HBs on the concentration of dipoles of the OH group ($C_{OH}$), the $C_{OH}$ value was calculated by dividing $\rho$ by the molecular weight of the liquid ($M_r$). For water, to take into account the vector sum of the dipoles of the two OH groups, the value of $C_{OH}$ was taken with the coefficient $2\cos(104.45/2)°$.

Empirical data for TDs characteristics of water, alcohols and hexane were imported from published sources. References to these materials are given in figure captions and in table. Graphs were digitalized with the use of 'Paint' computer application, when necessary. The borders of the temperature range for TDs were defined based on those of available data varying in the T limits of ~253 K to ~373 K. The points 273 K, 298 K, and 373 K were taken as the boundary points of the T-intervals TDs. 'MS Excel' application was used to plot TDs and their approximations. The extent of proximity of value $R^2$ to 1 was chosen as the reliability criterion, for approximations, in various T-intervals.

3. **Results**

Graphs for approximations and those for reference data are shown in Figures 1 to 10, while forms of functions $F_A$ and values of $E_A$ and $E_R$ are presented in the Tables 1 and 2.

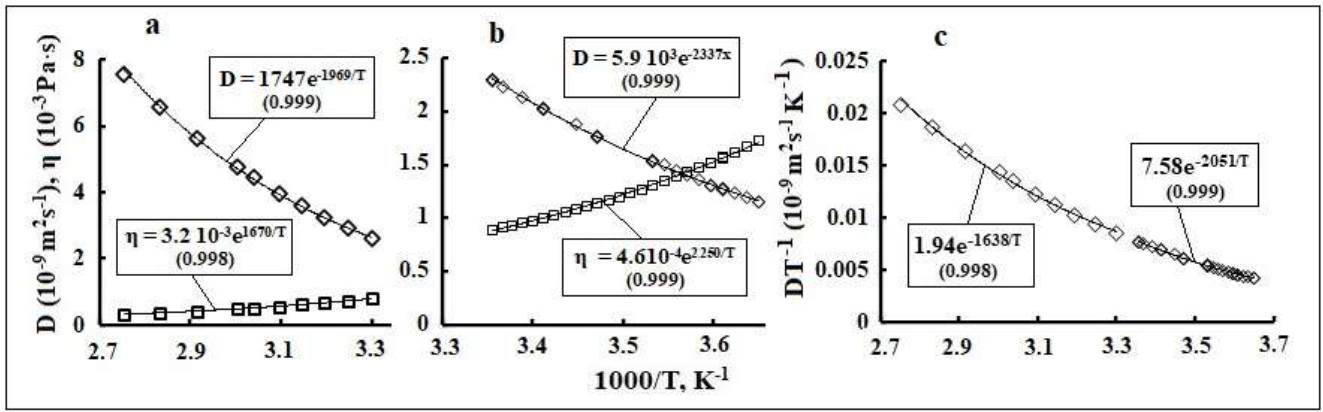

**Figure 1.** Dependences of viscosity ($\eta$), self-diffusion coefficient (D) and compositions $DT^{-1}$ of water on 1/T, and their Arrhenius approximations. Initial data for D from [22, 27, 29], for $\eta$ from [19, 35].

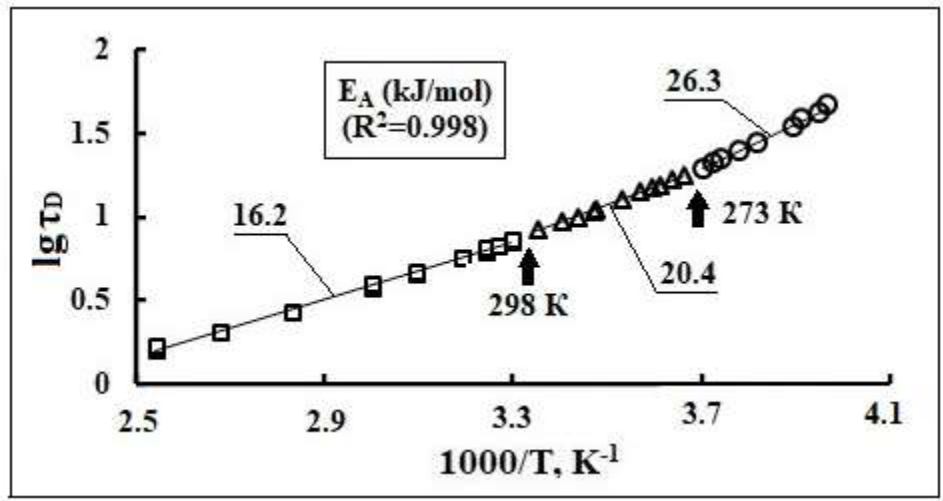

**Figure 2.** Dependences dielectric relaxation time ($\tau_D$) of water on 1/T, and its Arrhenius approximations. Initial data from [25].

Temperature interval, parameter β and activation energy of Arrhenius approximations of the temperature dependences of the water characteristics

| Water properties | β | ΔT (°C) | $E_A$ | $E_R$ | $E_T$ | [Ref] |
|---|---|---|---|---|---|---|
| | | | | kJ/mol | | |
| D ($cm^2\ s^{-1}$) | 1 | -20 – -1 | -26.6 | -24.6 | -2.0 | [25] |
| | | 0 – 25 | -19.4 | -17.0 | -2.4 | [22, 27 |
| | | 30 – 100 | -16.3 | -13.6 | -2.7 | 29, 32] |
| $\tau_D$ (ps) | -1 | -22 – -3 | 26.3 | 24.1 | 2.2 | |
| | | 0 – 25 | 20.4 | 18.0 | 2.4 | [25, 31] |
| | | 30 – 100 | 16.2 | 14.2 | 2.2 | |
| η (Pa s) | 0 | -20 – -1 | 23.5 | | 0 | |
| | | 0 – 25 | 18.7 | | 0 | [19, 35] |
| | | 26 – 100 | 13.9 | | 0 | |
| $\varepsilon_s$ | -1 | -20 – 0 | 2.7 | 0.5 | 2.2 | [20, 31, |
| | | 0 – 25 | 3.0 | 0.6 | 2.4 | 33, 34] |
| | | 25 – 100 | 4.3 | 1.5 | 2.8 | |

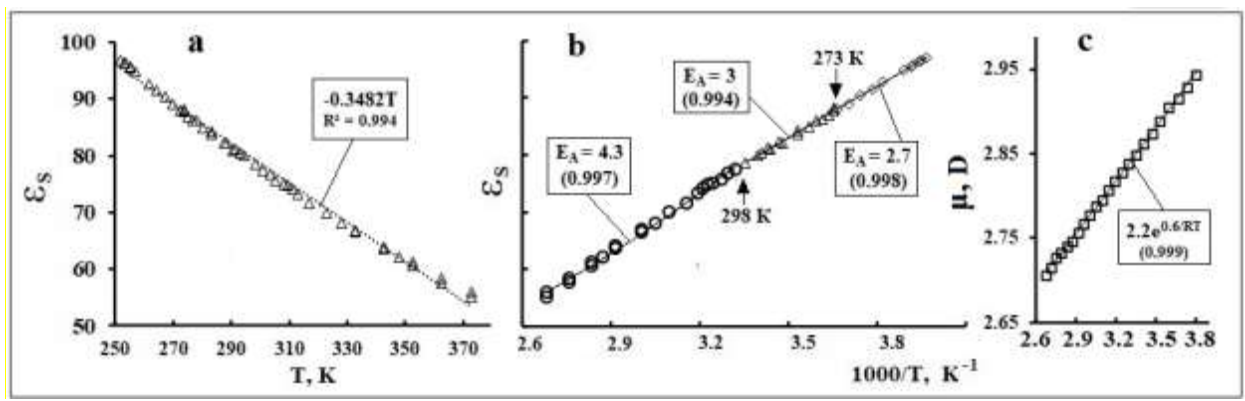

**Figure 3.** (**a**) Experimental dependence of dielectric constant ($\varepsilon_s$) on T and its linear approximation; (**b**) the dependence of $\varepsilon_s$ on 1/T and its $F_A$ approximation; Initial data from [20, 31]. (**c**) Dependence of the dipole moment of a water molecule (μ) on 1/T and its $F_A$ approximation. Initial data from [30]. $E_A$ in kJ/mol.

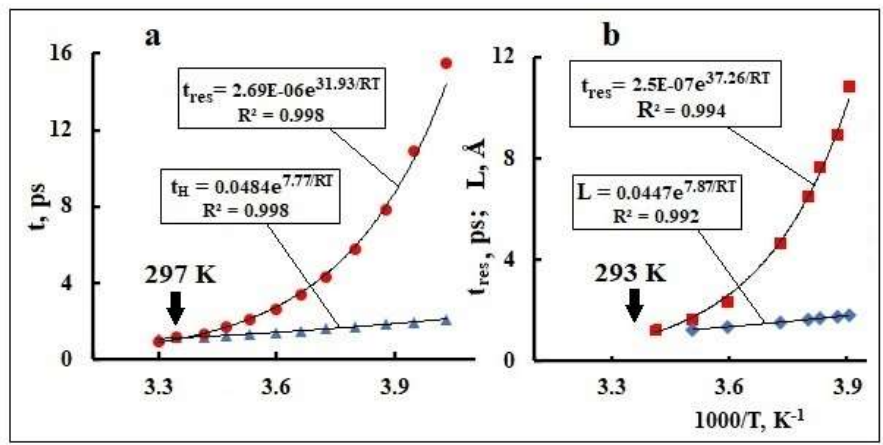

**Figure 4.** (**a**) Dependences of hydrogen bond lifetime ($t_H$) and calculated residence time ($t_{res}$) on 1/T and their $F_A$ approximation. Initial data from [36, 37]; (**b**) experimental dependence of $t_{res}$ and the mean jump length (L) on 1/T and their $F_A$ approximation. Initial data from [38, 39]. Arrows on (a) and (b) indicate the values of T at which $t_H$ and L are equal to $t_{res}$.

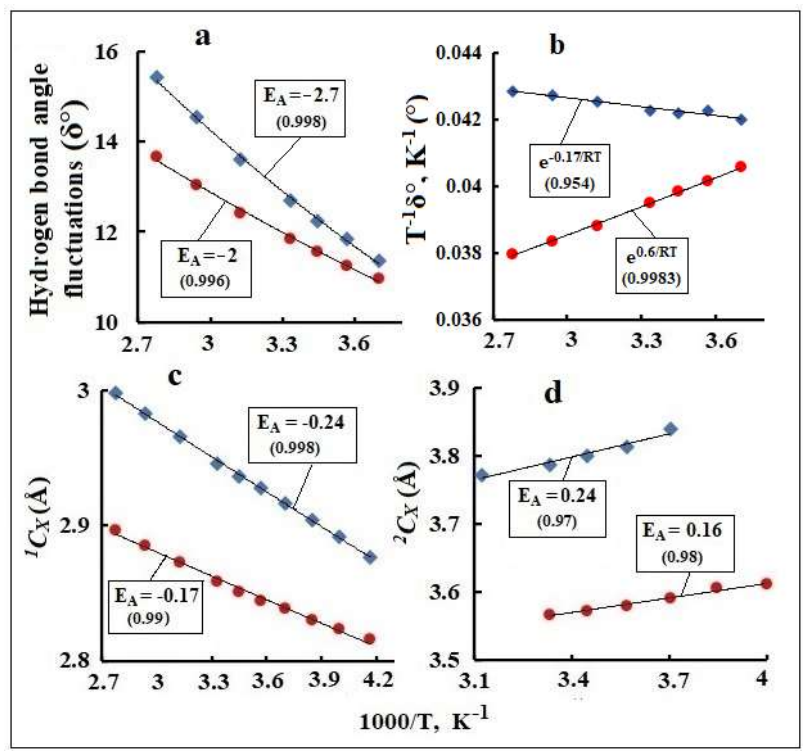

**Figure 5.** Dependences of the amplitude of fluctuations of the hydrogen bond angle ($\delta°$) and composition $T^{-1}\delta°$ on 1/T and their $F_A$ (**a**) and $F_R$ (**b**) approximations. Dependences of $g_{oo}(r)$ functions of the first ($^1Cx$) (**c**) and second ($^2Cx$) (**d**) shell on 1/T and their $F_A$-approximations. Activation energies $E_A$ and $E_R$ in kJ/mol. BLYP option (red circles) and RPBE option (blue diamond). See text. Initial data from [15].

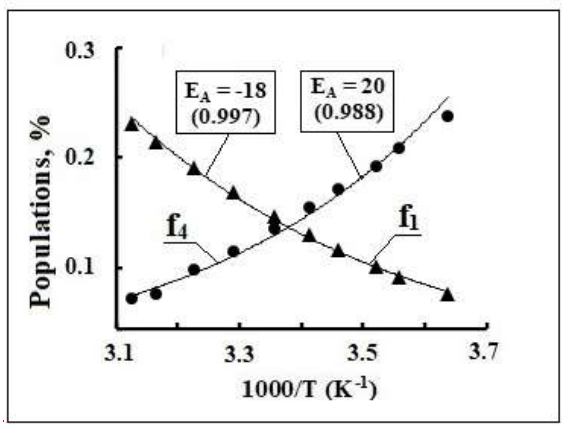

**Figure 6.** Temperature evolutions populations of molecules with O-H stretching modes $f_1$ (the shoulder 3620 cm$^{-1}$) and $f_4$ (the shoulder 3260 cm$^{-1}$), and their $F_A$ approximations. Initial data from [40, 41].

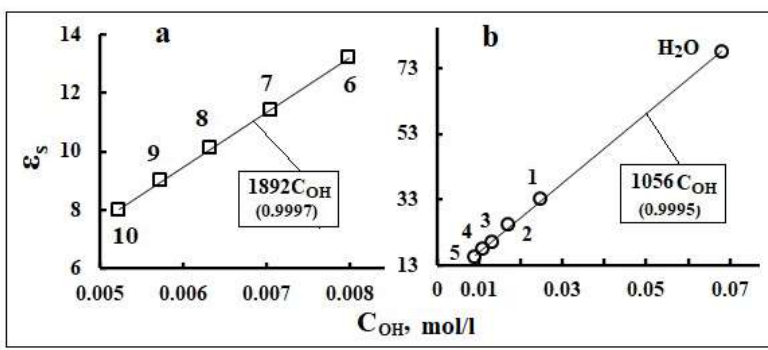

**Figure 7.** Linear trends in the dependences of the dielectric constants ($\varepsilon_s$) at 298 K of water and a number of alcohols (numbers correspond to N from 1 to 10 in Table 2) on the concentration of the dipole moments of the OH group ($C_{OH}$). Initial data for (**a**) from [42] and for (**b**) from [43, 44].

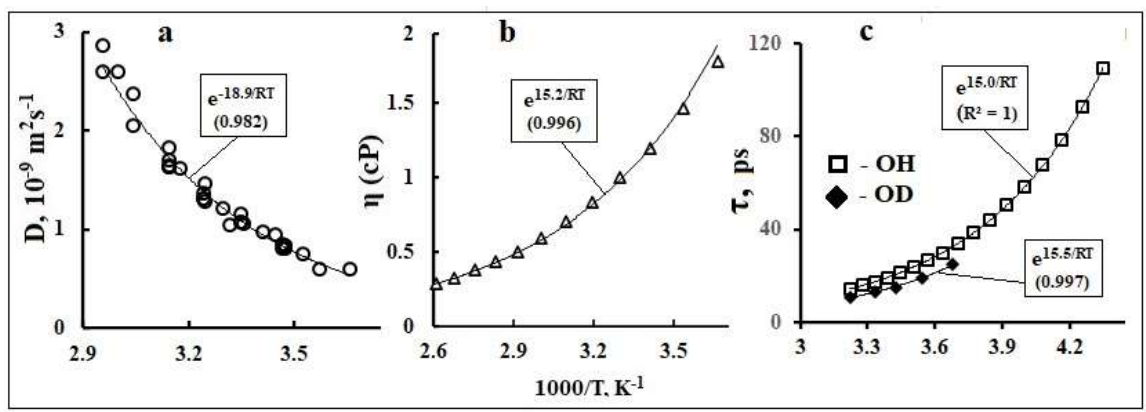

**Figure 8.** Dependences on 1/T of ethanol self-diffusion coefficient (D) (**a**); viscosity ($\eta$) (**b**) and reorientation correlation times for $CD_3CD_2OH$ and $CH_3CH_2OD$ ($\tau$, ps) (**c**). In brackets are $R^2$, $E_A$ values in exponents in kJ/mol. Initial data from [27, 45-48].

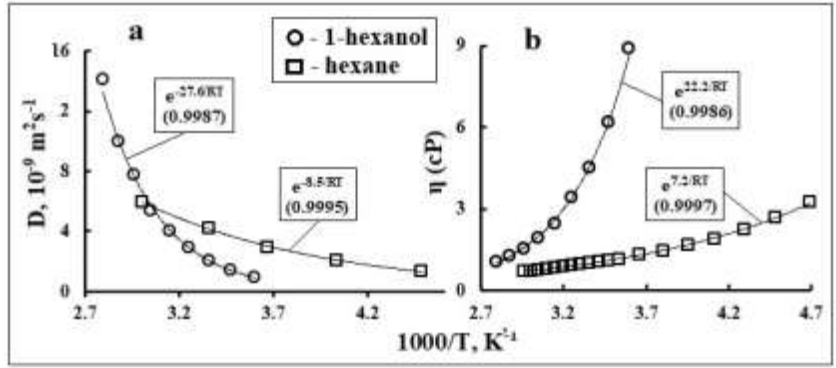

**Figure 9.** Dependences on 1/T of self-diffusion coefficient (D) (**a**) and viscosity (η) (**b**) 1-hexanol and hexane, and their Arrhenius approximations. $E_A$ values in exponents in kJ/mol. Initial data from [42, 49].

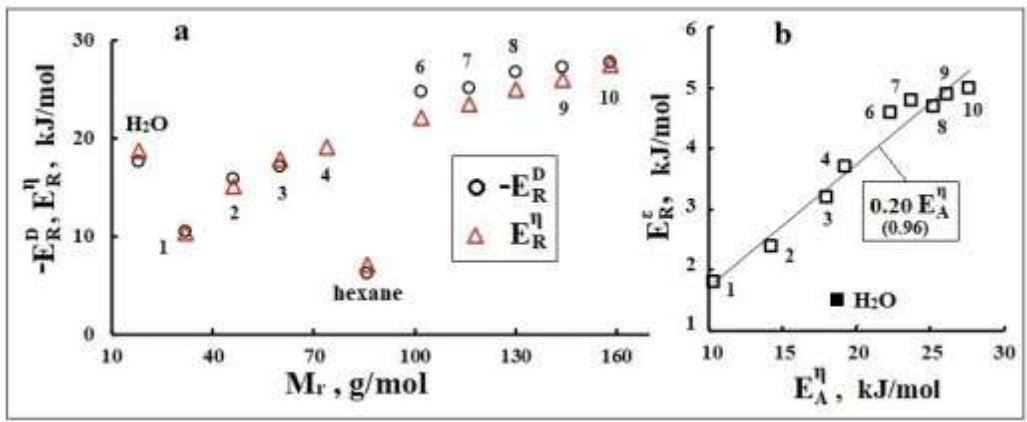

**Figure 10.** (**a**) Dependence of the activation energies ($E_R$) of the reactions of rearrangement of the structure of water, alcohols and hexane on the molecular weight ($M_r$). Dependences of the activation energy of the dielectric constant ($E_R^\varepsilon$) on the activation energies of the viscosity ($E_a^\eta$) for alcohols and water. N points of alcohols from 1 to 10 in Table 2. Initial data from [27, 42, 43, 47, 48, 49].

**Table 2**

Temperature interval and activation energy of Arrhenius approximations of temperature dependences of alcohols and hexane dynamic characteristics

| N | Fluid | | ΔT (°C) | E (kJ/mol) | | | | | [Ref] |
|---|---|---|---|---|---|---|---|---|---|
| | | | | $-E_A^D$ | $-E_R^D$ | $E_A^\eta = E_R^\eta$ | $E_A^\varepsilon$ | $E_R^\varepsilon$ | |
| 1 | | methanol | -2 – 76 | 13.0 | 10.5 | 10.4 | 4.3-4.4 | 1.8 | [27, 43, 47, 48, 49] |
| 2 | | ethanol | 0 – 65 | 18.3 | 15.9 | 15.2 | 4.8 | 2.4 | |
| 3 | | propanol | 0 – 55 | 19.8 | 17.2 | 17.9 | 5.8 | 3.2 | |
| 4 | | butanol | 0 – 60 | - | - | 19.2 | 6.2 | 3.7 | |
| 5 | 1- | pentanol | 0 – 50 | 24.1 | - | - | 5.8 | - | [29, 43] |
| 6 | | hexanol | | 27.5 | 24.8 | 22.2 | 7.3 | 4.6 | |

| 7 | heptanol | 15 – 85 | 27.9 | 25.2 | 23.6 | 7.5 | 4.8 | |
| --- | --- | --- | --- | --- | --- | --- | --- | --- |
| 8 | octanol | | 29.5 | 26.8 | 25.1 | 7.4 | 4.7 | [42] |
| 9 | nonanol | | 30 | 27.3 | 26 | 7.6 | 4.9 | |
| 10 | decanol | | 30.5 | 27.8 | 27.5 | 7.7 | 5 | |
| 11 | hexane | 10 – 65 | 8.5 | 6.3 | 7.2 | 0.4 | -1.7 | [49-51] |

The ARA trends of temperature dependences of D, η, $τ_D$ and $ε_s$ of water have kinks at points 273 K (except for η) and 298 K (see Table 1, Figures 1 and 2). The moduli of the $E_R$ values for D, η and $τ_D$ when the water is cooled below 298 K increase by ~35%, and for $ε_s$, on the contrary, decreases by ~60%. In this case, the $E_R$ for $ε_s$ and μ in the range 273-298 K are close and 30 times less than $E_R$ for D, η, $τ_D$. For D, η, $τ_D$ alcohols, and even more so hexane, jumps in $E_A$ and $E_R$ values in the vicinity of 298 K are not detected within the measurement errors (see Table 2, Figures 8 and 9). From the dependence of $E_R^D$ and $E_R^η$ values on $M_r$ for alcohols drops out points for water and hexane (Figure 10 a), and their values for 1-hexanol are three times higher than for hexane. Hence, it follows that the mobility of hydrogen-bonded liquids is mainly determined by the HBs density. In alcohols, the number of HBs per molecule can reach three, while the difference in $E_A^η$ for 1-hexanol and hexane, equal to 15 kJ/mol, leads to the estimate $E_H$ = 5 kJ/mol. For water, the HB number per molecule in the range of 273-298 K is ~3.5 [5], therefore, at $E_A^η$=18.8 kJ/mol (see Table 1), the average $E_H$ value is also close to 5 kJ/mol.

The $ε_s$ value of water is ~2.5 times greater than the $ε_s$ of alcohols (see Figure 7), and μ of water is enhanced to 2.6 D from 1.84 D in the gas phase [30, 52-55]. A significant decrease in $E_R^ε$ for water compared to $E_R^ε$ for alcohols (Figure 10 b) indicates a high contribution of water SMS domains to the dependence of the volume polarization on T. Differences in the dependences of the rotational and translational diffusion of water and alcohols on $M_r$ and η are due to the small size water molecule, a high concentration of HBs (Figures 7 and 10) and the dominance of clusters with tetrahedral HBs in its SMS [1, 5, 15-18]. Taking into account the significant differences between TDs $ε_s$ and the dynamic characteristics of water, the $ε_s$ parameter can serve as an adequate marker of the effects of intermolecular interactions. In contrast to $ε_s$, the kinetic parameter $τ_D$, like $τ_r$, characterizes the MD of water only at the level of an individual molecule and its immediate environment [25, 26, 28, 31, 41, 52-56].

**4. Discussion**

**4.1. Structure and molecular dynamics of water in range 273-298 K**

The jump in the values of $E_A$ and $E_R$ at the point 298 K for $ε_s$ and D, η, $τ_D$ of water (see Figures 1, 2 and 3) and the deviation of TDs of these characteristics from Arrhenius [19-24, 36] are a consequence of the dynamic phase transition in the vicinity of 298 K [10]. This hypothesis is consistent with the representation of the structure of liquid water by the «ice-like framework» model [57]. This model in the form of a metastable «crystal-like» structure was used to explain the features of the thermodynamics of aqueous solutions of biological substances at T from 273 K to ~300 K in [10]. Computer simulations of the structure of supercooled water [58] confirm the presence in the ice-like model of modifications of ice structures Ih and Ic in the form of alternating mirror-symmetric layers of six-membered deformed rings.

Similar ice-like SMS are formed upon melting ice 273 K in the form of the following hexamers configurations: cyclic, cage, prism, book [15, 59]. The energies of the ground states of the three-dimensional cage and prism isomers differ by ~1 kJ/mol, and their precursor is the cyclic isomer, which appears immediately after the melting of ice [15, 57-59]. The $\mu$ values in the series cyclic, cage and prism isomer are equal to 0 D; 1.9 D; 2.7 D [53], respectively. A zero value for $\mu$ for cyclic hexamer indicates a pairwise combination of dipoles in a ring with opposite $\mu$ directions. The similarity of the structure of the metastable phase of SMS water in the supercooled state at T<273 K and ordinary water in the range of 273-298 K is evidenced by the fact that the $E_A$ and $E_R$ values for $\varepsilon_s$ before and after 273 K differ by ~10% and 20%, and before and after 298 K by ~33% and ~115%, respectively (see Figure 3, Table 1).

The persistence of the effect of the ice-like phase on the MD of water up to 298 K is confirmed by reliable extrapolations of the dependences of $t_{res}$, L and $t_H$ on 1/T from the range ~250-273 K to the range 273-298 K (Figure 4). In papers [35, 36, 38, 39, 56, 60], it is also believed that the excess of $t_{res}$ over $t_H$ and over the mean jump time of a molecule ($t_j$) is due to a significant contribution to the kinetics of self-diffusion of water from the MD ice-like phase. The value of $t_j$ is taken equal to L/$v$, where $v$ is the average velocity of the translational motion of molecules. The conditional boundary temperature for the existence of the ice-like phase is T, at which $t_{res}$ becomes greater than the values of $t_H$ and $t_j$. This requirement is fulfilled after the intersection point of the $F_A$ approximations TDs $t_{res}$, $t_H$ and $t_j$. The obtained values of 297 K and 293 K differ by only 1% - 2% from 298 K (see Figure 4).

MD simulation [60] showed that at 298 K the maximum concentration of interstitial molecules is observed and this leads to anomalies in the distribution functions and thermodynamic properties of water. The appearance of interstitial molecules at 298 K can be associated with the decay of the metastable ice-like phase. Note that the dynamics of interstitial molecules plays the same role in the TDs of distributions of molecules in the first and second shells of the central molecule [15].

### 4.2. Resonant mechanism of hydrogen bond rearrangement

Reversible phase transitions at 273 K and 298 K occur simultaneously and cover the entire volume of water. Apparently, their thermodynamics are based on resonant mechanisms of absorption and reemission of thermal energy quanta [1] by coherent systems of HBs in ice-like structure of a metastable water phase [57]. It follows from MD simulation [61] that when an aqueous solution of $CO_2$ is heated in the range of 275–300 K, jumps in the internal energy are observed as during a phase transition, while the rearrangement of the short-range order of all molecules occurs in a time less than ~1 ns. Structural analysis of neighboring pairs of water molecules showed that after melting of the ice-like structure, the number of HBs decreases by ~10-15% while maintaining the integrity of a single network of HBs.

In [62], a formula was obtained for the frequency of coherent vibrations of *n* protons in a spiral chain of tetrahedral HBs:

$$\nu = 22n^{-1}(n-1)^{1/2} \text{ (THz)}.$$

From this formula, for n=12, we obtain $\nu$ = 6.08 THz and the vibration energy of ~2.4 kJ/mol, which is equal to the thermal energy at T = 298 K. Since the main element of the ice-like structure is tetrahedral clusters hexamers containing 12 protons, then it can be assumed that at the point 298 K there is a resonant rearrangement of hexamers into helical chains of 6 water molecules, followed by their decay into dimers and interstitial molecules. By the same mechanism, the structure of water can be rearranged at the extrema TDs ($T_E$) of ρ and adiabatic heat capacity at 277 K and 308 K, respectively, since at these points the relation $E_R \approx RT_E$ is fulfilled [18].

The possibility of resonant transfer of vibrational-rotational energy over the HBs network in the vicinity of 298 K is evidenced by the following. Water absorption band at ~170-180 $cm^{-1}$ (~5-6 THz) refers to the intermolecular vibration of the HB network with relatively strong contribution of the HB stretch vibration [1, 33, 34]. These vibrations can have out-of-plane polarization and are associated with fluctuations of the charge along the HB [40, 63, 64]. When water is irradiated with THz pulses in this absorption band, ~85% of the external energy turns into excitation of resonant collective rotational degrees of freedom in the tetrahedral HBs network, and 15% turns into the translation of molecules [1]. Under optical excitation of stretch vibration of O–H bond (~3400 $cm^{-1}$) of water, there is a fast resonant intermolecular transfer of vibrational energy from the OH bond to many molecules, before the excitation energy dissipates into low-frequency vibrations of the network of HBs [65-68]. In the transfer mechanism, along with dipole-dipole interactions of O-H bonds, intermolecular anharmonic interactions hydrogen bonds can be involved [67]. In alcohols, the processes of relaxation of external energy follow similar mechanisms, but the relaxation times of the network obtained f HBs increase with the lengthening of the alkyl chain [66, 69].

### 4.3. The hydrogen bond fluctuations in water

Rotational diffusion of molecules, intermolecular Coulomb interactions, and asymmetry of the charge distribution in HBs are manifested by fluctuations of the HB angle (δ°) relative to the average symmetric tetrahedral HBs configuration on a scale of hundreds of femtoseconds [15, 70]. Averaged over two variants of calculations (see Figure 5), the values of <$E_A$> and <$E_R$> for $\delta°$ are equal to -2.36 kJ/mol and 0.2 kJ/mol, and <$E_A$> of the reactions of rearrangements of configurational shells $^1$Cx and $^2$Cx are equal to -0.2 kJ/mol and 0.2 kJ/mol, respectively. The value <$E_A$> = -2.36 kJ/mol for $\delta°$ is equal to the thermal energy at the midpoint of the interval 273-298 K, which allows the resonant mechanism of excitation of fluctuations of the angle HB. This assumption is consistent with the equality <$E_R$> for $\delta°$ and <$E_A$> for the exothermic reaction of rearrangement of the $^2$Cx shell. The average values of fluctuation amplitude $\delta°$ were 11.1° ± 0.2° at 273 K and 12.2° ± 0.4° at 298 K. Both values are 10-12% of the equilibrium angle of ~104° in the tetrahedral HBs system. According to empirical Lindemann melting rule, the critical share of the amplitude of thermal vibrations varies in the range of 5-15% [71]. *Ab initio* MD simulations based on the partitioning of the total interaction energy of molecules [70] have shown that the strength and configuration of HBs change considerably in $\delta°$ range ~5° to 10°. Therefore, fluctuations of HBs at 273 K can serve as a factor triggering the melting of ice Ih, and at 298 K the decay of the metastable ice-like phase.

### 4.4. Energy of dynamics of hydrogen bonding network

In the range 273-298 K, the thermal energy (RT) is of the same order of magnitude as the stretch vibration energy HBs (~180 cm$^{-1}$), as well as the $E_A$ values of HBs fluctuations and transformations of polarized domains in the HBs network of water. The $E_A$ values for D, $\eta$, $\tau_D$ and stretch vibration of the O-H group are an order of magnitude higher than RT (see Table 1, Figures 5 and 6). Accordingly, the MD kinetics is divided into two components – fast vibrational-translational motions of individual molecules and slow diffusion of thermal energy over the degrees of freedom of the HBs network [65–68]. At constant temperature and in the absence of external influence, the kinetics of MD is limited by the mechanisms of generation and accumulation of thermal energy quanta, both on an individual molecule and on domains from polarized dipoles. These mechanisms are based on electrical interactions of fluctuating HBs dipoles, allowing the formation of correlated ensembles of HBs and resonant energy exchange between them [1, 15, 28, 40, 62, 63, 72-76]. The participation of thermal energy in the dynamics of these processes adequately reflects the component $T^{\pm\beta}$ in the approximations of TDs (5). At the same time, the signs of $\beta$ and $E_T$ correlate with the thermal effects of endothermic (for D and $\delta°$) and exothermic (for $\tau_D$ and $\varepsilon_s$) rearrangements in the structure of water (see Figure 11).

The average number of HBs in tetrahedral configurations of the HBs network in the range of 273–298 K fluctuates in the range from 3 to 5 [5, 74, 76]. As a result, the HBs network structure always contains molecules with vacancies for donor and acceptor HBs and irregularities in charge distributions in the volume of water [73–77]. The Coulomb interactions between them initiate energetically consistent rearrangements of HBs within the first two shells of the central molecule [15, 59, 75, 76, 78, 80]. This is evidenced by the opposite signs and close $E_A$ values of synchronous $^1Cx$ and $^2Cx$ changes and HBs rearrangements in populations $f_1$ and $f_4$ (see Figures 5 and 6). The synergism of exothermic reactions of formation and endothermic reactions of HBs breaking provides activation of the water molecule reorientation and a consistent translational jump of the molecule outside the cell [78-82]. The relationship and energy balance of the motions of molecules responsible for viscosity and self-diffusion is confirmed by the dependence $E_R^\varepsilon$ on $E_a^\eta$ and the closeness of the moduli of the $E_R$ values for $\eta$ and D (see Figure 10 and 11).

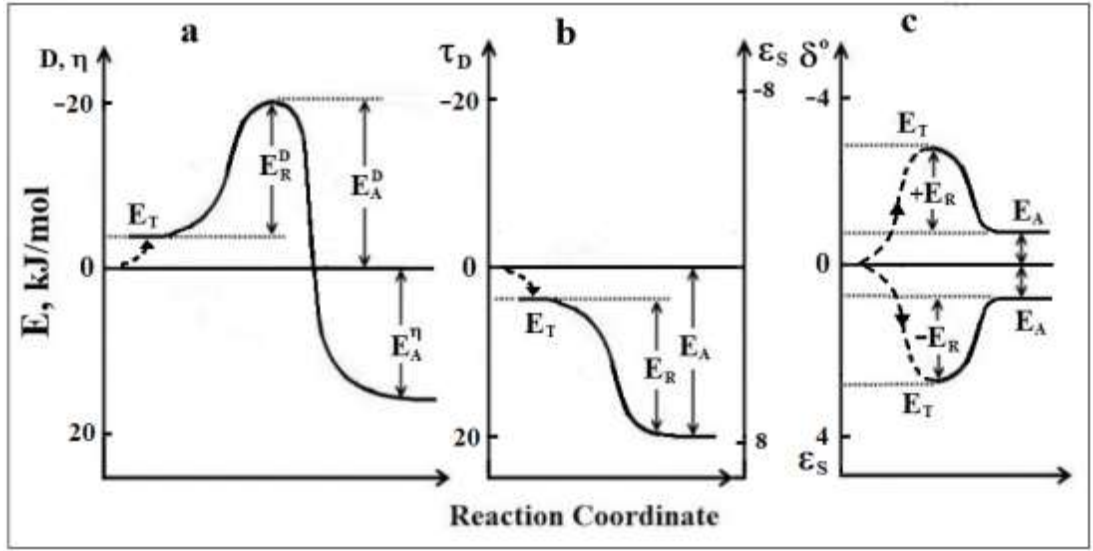

**Figure 11.** Schemes of energy profiles of processes in the range of 273-298 K of self-diffusion (D) and viscosity ($\eta$) of water and alcohols (**a**), their dielectric constant ($\varepsilon_s$) and dielectric relaxation ($\tau_D$) of water (**b**), fluctuations of hydrogen bonds of water ($\delta°$) and dielectric constant of hexane (**c**). Dotted arrows - actions of thermal energy. See Tables 1 and 2 for exact activation energies ($E_R$, $E_R$, and $E_T$).

Relations (3) are extrapolations to the molecular level TD of the SE equation [21]:

$$D = \frac{k_B T}{6\pi\eta r}, \qquad (6)$$

in which D is the diffusion coefficient of a Brownian particle of radius r in a liquid with viscosity $\eta$ and $k_B$ is the Boltzmann constant (1.38 $10^{-23}$ J K$^{-1}$). For self-diffusion of a liquid, r means the conditional «molecular radius» [21]. In the case of water, a significant contribution of HBs to MD leads to a complication of the relationships between the micro parameters r, $\eta$, D, which leads to distortions of equation (6). In this case, modifications (6) by fractional equations (3) do not take into account the possibility of TD of r, due to the clustering of the structure of water and especially at T <298 K. The formal expression of the synergism of molecular physics of self-diffusion and viscosity is the product D$\eta$, which has the dimension of force (N). The value of D$\eta$ adequately reflects the nature of the friction force in a liquid, summing up the effect of its nearest environment on the rotation and translation of a molecule [82]. In this context, relation (6) is transformed to the form:

$$D\eta r \propto kT,$$

which formally expresses the dependence of the work of friction forces at a distance of the order of the particle radius on thermal energy.

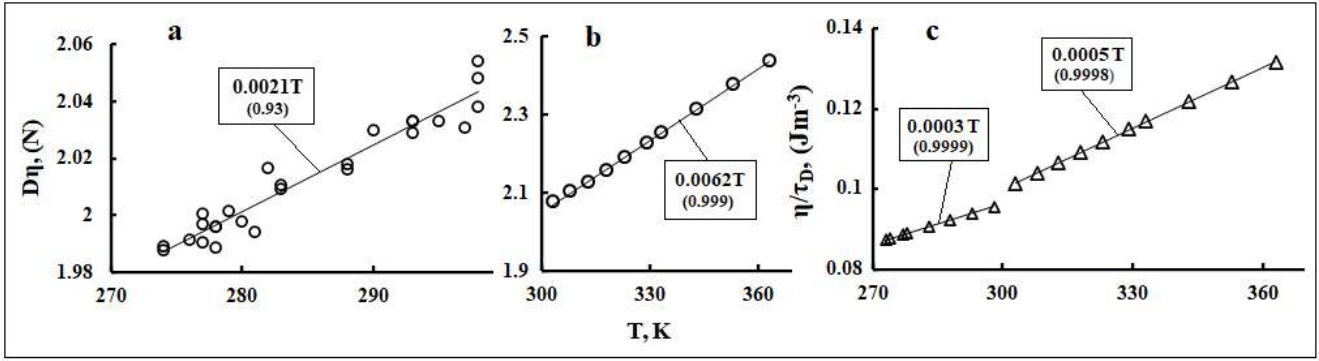

**Figure 12.** Linear trends in the dependences on T of compositions $D\eta$ (**a, b**) and $\eta/\tau_D$ (**c**). Initial data from Figure 1.

Using the data for D and $\eta$ (see Figure 1), we built the dependence of $D\eta$ on T (see Figure 12) and, taking into account (6), calculated the value of r from the coefficient of the linear trend. For the ranges 273-298 K and 300-363 K, the r values were 3.5 Å and 1.2 Å, respectively. The value r ~3.5 Å is typical for tetrahedral hexamers and domains formed from molecules with strongly correlated μ [59, 72, 75, 76]. Apparently, these formations limit the mobility of the ice-like water phase in the range of 273–298 K. The value of r ~ 1.2 Å correlates with the length of the O–H bond (0.96 Å) and the polarizability of the water molecule in the gas phase (1.44 Å³) [75] . For ethanol from the data in Figure 7, an estimate of r ~ 1.1 Å is also obtained in the entire range from 273 K to 340 K. It follows that the mobility of water in the range of 298-373 K and ethanol in the range of 273-340 K is mainly responsible for «argon-like» dynamics of almost free molecules [10].

If we separate the thermal component from TD of D in accordance with eqution (5) and Figure 1c, we can obtain physically adequate analogs of formulas (3) for two ranges of T from the ARA presented in Figure 1a and 1b:

$$D\ (273\text{-}298\ K) = 3.5\ 10^{-15} \frac{T}{\eta} \exp\left(\frac{1.7}{RT}\right);$$

$$D\ (300\text{-}363\ K) = 6.2\ 10^{-15} \frac{T}{\eta} \exp\left(\frac{0.26}{RT}\right)$$

Dimension D and $\eta$ in SI system, and $E_R$ value in exponentials in kJ/mol.

By analogy with (4), SED is transformed to the form [21, 25]:

$$1/\tau_D = \frac{k_B T}{4\pi\eta r^3}. \qquad (7)$$

Taking into account (7), the r values were estimated from the linear trends of the TD composition η/τ$_D$ (see Figure 2). For the ranges 273-298 K and 300-363 K, r values of 1.5 Å and 1.3 Å were obtained, respectively. These values are close to the recommended value of the distance between molecules 1.44 Å at 0 °C [25], but to a lesser extent depend on T than r determined from TD of Dη. This may be due to the method of measuring τ$_D$ from the relaxation time of individual molecules, resonantly excited by external energy.

## 5. Conclusion

The study of the dependence of the dynamic and dielectric properties of water, alcohols and hexane on temperature in the range of 273-298 K using modified Arrhenius approximations showed that the thermodynamics of phase transitions and anomalies of water properties are based on fluctuations of hydrogen bond dipoles and long-range interactions between them. The close-range interactions of charges on molecules and dipoles of hydrogen bonds, as well as the synergism of endothermic and exothermic reactions of their formation and rupture, determine the features of the energetics of the molecular dynamics of water and alcohols. Correlations of fluctuations of hydrogen bonds and the resonant mechanism of their excitation by thermal energy are responsible for the anomalies of the temperature dependences of water properties and for phase transition of ice Ih into the metastable ice-like phases of water at 273 K, as well as for the decay of the ice-like phase in the vicinity of 298 K.

**References**

1. H. Elgabarty, et al., Energy transfer within the hydrogen bonding network of water following resonant terahertz excitation, Science Advances. 6 (2020) 7074: 10.1126/sciadv.aay7074
2. L. M. Pettersson, Y. Harada, A. Nilsson, Do X-ray spectroscopies provide evidence for continuous distribution models of water at ambient conditions? PNAS. 116 (2019) 17156-57; https://doi.org/10.1073/pnas.1905756116 /.
3. J. Niskanena, et al., Reply to Pettersson et al.: Why X-ray spectral features are compatible to continuous distribution models in ambient water, PNAS. 116 (2019) 17158-59; https://doi.org/10.1073/pnas.1909551116
4. L.G.M. Pettersson, R.H. Henchman, A. Nilsson, Water – The Most Anomalous Liquid, Chem. Rev. 116 (2016) 7459-62, https://doi.org/10.1021/acs.chemrev.6b00363.
5. M.F. Chaplin, Water Structure and Science: http://www1.lsbu.ac.uk/water/index.html.
6. A. Kholmanskiy, Chirality anomalies of water solutions of saccharides. J. Mol. Liq. (2016) 216, 683-7. 10.1016/j.molliq.2016.02.006
7. B. Bagchi, Water in biological and chemical processes: from structure and dynamics to function. 2013.
8. L.G.M. Pettersson, A Two-State Picture of Water and the Funnel of Life, (2019) In book: Modern Problems of the Physics of Liquid Systems, 3-39: 10.1007/978-3-030-21755-6_1.
9. A.S. Kholmanskiy, Dependence of Wheat Seed Germination Kinetics on Temperature and Magnetic Field. Res. J. Seed Sci. 9 (2016) 22-28. 10.3923/rjss.2016.22.28
10. V. Bardik, et al., The crucial role of water in the formation of the physiological temperature range for warm-blooded organisms, J. Mol. Liq. 306 (2020) 112818, https://doi.org/10.1016/j.molliq.2020.112818
11. M.D. Ediger, Spatially heterogeneous dynamics in supercooled liquids, Annual Rev. Phys. Chem. 51 (2000) 99-128. 10.1146/annurev.physchem.51.1.99

12. G.P. Johari, J. Teixeira, Thermodynamic analysis of the two-liquid model for anomalies of water, J. Phys. Chem. B. 119 (2015) 14210. 10.1021/acs.jpcb.5b06458
13. J. Niskanena, et al., Compatibility of quantitative X-ray spectroscopy with continuous distribution models of water at ambient conditions, PNAS. 116 (2019) 4058-63. 10.1073/pnas.1815701116.
14. V. Vaz da Cruz, et al., Probing hydrogen bond strength in liquid water by resonant inelastic X-ray scattering. Nat. Commun. 10 (2019), 1013 https://doi.org/10.1038/s41467-019-08979-4
15. T. Morawietz, A. Singraber, C. Dellago, J. Behler, How van der Waals interactions determine the unique properties of water, PNAS. 113 (2016) 8368-73. https://doi.org/10.1073/pnas.160237511321.
16. A. Kholmanskiy, The supramolecular physics of the ambient water, (2020): arXiv:1912.12691v1 .
17. A. Kholmanskiy, Activation energy of water structural transitions, J. Mol. Struct. 1089 (2015) 124-128. https://doi.org/10.1016/j.molstruc.2015.02.04919
18. A. Kholmanskiy, N. Zaytseva, Physically adequate approximations for abnormal temperature dependences of water characteristics, J. Mol. Liq. 275 (2019) 741-8 https://doi.org/10.1016/j.molliq.2018.11.059
19. J. Pátek, et al., Reference Correlations for Thermophysical Properties of Liquid Water at 0.1 MPa, J. Phys. Chem. Ref. Data. **38**, 21 (2009); https://doi.org/10.1063/1.3043575
20. W. Ellison, Permittivity of pure water, at standard atmospheric pressure, over the frequency range 0–25 THz and the temperature range 0–100 °C. J. Phys. Chem. Reference Data (2007). 10.1063/1.2360986
21. M. G. Mazza, et al., Connection of translational and rotational dynamical heterogeneities with the breakdown of the Stokes-Einstein and Stokes-Einstein-Debye relations in water. Phys. Rev. E 76 (2007) 031203. 10.1103/PhysRevE.76.031203
22. J.C. Hindman, Relaxation processes in water: Viscosity, self-diffusion, and spin-lattice relaxation. A. kinetic model. J. Phys. Chem. 60 (1974) 4488–93.
23. K.R. Harris, Communications: The fractional Stokes–Einstein equation: Application to water. J. Chem. Phys. 132 (2010) 231103. DOI: 10.1063/1.3455342
24. I. N. Tsimpanogiannis, et al., (2020). On the validity of the Stokes–Einstein relation for various water force fields. Mol. Phys. [e1702729]. https://doi.org/10.1080/00268976.2019.1702729
25. N. Agmon, Tetrahedral Displacement: The Molecular Mechanism behind the Debye Relaxation in Water J. Phys. Chem. 100 (1996) 1072-1080. 10.1021/jp9516295
26. K.Winkler, et al., Ultrafast Raman-induced Kerr effect of water: Single molecule versus collective motions. J. Chem. Phys. 113 (2000) 4674; http://dx.doi.org/10.1063/1.1288690
27. G. Guevara, J. Vrabec, H. Hasse, Prediction of self-diffusion coefficient and shear viscosity of water and its binary mixtures with methanol and ethanol by molecular simulation, J. Chem. Phys. 134 (2011) 074508. 10.1063/1.3515262
28. R. A. Nicodemus, et al., Collective hydrogen bond reorganization in water studied with temperature-dependent ultrafast infrared spectroscopy, J. Phys. Chem. B. 115 (2011) 5604-16. 10.1021/jp111434u
29. M. Holz, S.R. Heil, A. Sacco, Temperature-dependent self-diffusion coefficients of water and six selected molecular liquids for calibration in accurate $^1$H NMR PFG measurements, Phys. Chem. Chem. Phys. 2 (2000) 4740–5. DOI:10.1039/B005319H
30. A.V. Gubskaya, P.G. Kusalika, The total molecular dipole moment for liquid water, J. Chem. Phys. 117 (2002) 5290. https://doi.org/10.1063/1.1501122
31. D.C. Elton, The origin of the Debye relaxation in liquid water and fitting the high frequency excess response. Phys. Chem. Chem. Phys. 19 (2017) 10.1039/C7CP02884A.
32. O. Dietrich, Diffusion Coefficients of Water: 2002, https://dtrx.de/od/diff/
33. A.A. Volkov, V. Artemov, N.N. Sysoev, Possible mechanism of molecular motion in liquid water from dielectric spectroscopy data, J. Mol. Liq. 248 (2017) 564-568. 10.1016/j.molliq.2017.10.071
34. I. Popov, P. Ben Ishai, A. Khamzin, Y.D. Feldman, The mechanism of the dielectric relaxation in water. Phys Chem Chem Phys. 18 (2016). 10.1039/C6CP02195F
35. S. Dueby, V. Dubey, S. Daschakraborty, Decoupling of self-diffusion from viscosity of supercooled water: role of translational jump-diffusion, J. Phys. Chem. B. 123 (2019) 7178. 10.1021/acs.jpcb.9b01719
36. J. Teixeira, A. Luzar, Physics of Liquid Water. Structure and Dynamics, in "Hydration Processes in Biology: Theoretical and experimental approaches", NATO ASI series, Amsterdam, 1999.
37. J. Teixeira. The contribution of small angle and quasi-elastic scattering to the physics of liquid water. J. Phys. Conference Series. 848 (2017) 012003. https://doi.org/10.1088/1742-6596/848/1/012003

38. J. Teixeira, et al., Experimental determination of the nature of diffusive motions of water molecules at low temperatures. Phys. Rev. A 31 1985.  10.1103/PhysRevA.31.1913
39. J. Teixeira, A. Luzar and S. Longeville, Dynamics of hydrogen bonds: How to probe their role in the unusual properties of liquid water, J. Phys.: Cond. Matter 18 (2006) S2353. 10.1088/0953-8984/18/36/S09
40. J.D. Smith, et al. Unified description of temperature-dependent hydrogen-bond rearrangements in liquid water. PNAS, 102 (2005) 14171. https://doi.org/10.1073/pnas.0506899102
41. J-B. Brubach, et al., Signatures of the hydrogen bonding in the infrared bands of water. Chem. Phys. 122 (2005) 184509. 10.1063/1.1894929
42. A.M. Fleshman, et al., Describing temperature-dependent self-diffusion coefficients and fluidity in 1- and 3- alcohols using the compensated Arrhenius formalism, J. Phys. Chem. B, 120 (2016) 9959–9968. https://doi.org/10.1021/acs.jpcb.6b03573
43. R.D. Bezman, E.F. Casassa, R. L. Kay, The temperature dependence of the dielectric constants of alkanols, J. Mol. Liq. 73,74 (1997) 397-402
44. Engineering ToolBox, Dielectric Constants of Liquids.  https://www.engineeringtoolbox.com/liquid-dielectric-constants-d_1263.html.
45. S. Pothoczki. L. Pusztai, I. Bakó, Temperature dependent dynamics in water-ethanol liquid mixtures J. Mol. Liq. 271 (2018): 10.1016/j.molliq.2018.09.027
46. A. Klinov, I. Anashkin, Diffusion in Binary Aqueous Solutions of Alcohols by Molecular Simulation. Processes 7(12) (2019) 947; https://doi.org/10.3390/pr7120947
47. O. Suárez-Iglesias, et al., Self-Diffusion in Molecular Fluids and Noble Gases: Available Data J. Chem. Engineering Data 2015 60 (10), 2757-2817. https://doi.org/10.1021/acs.jced.5b00323
48. M. J. Assael, S.K. Polimatidou, Measurements of the viscosity of alcohols in the temperature range 290-340 K at pressures up to 30 MPa. Int. J. Thermophys. 15 (1994) 95-107. 10.1007/BF01439248
49. Dortmund Data Bank Dielectric Constant of 1-Butanol. http://www.ddbst.de/en/EED/PCP/DEC_C39.php
50. K.R. Harris, Temperature and Density Dependence of the Self-diffusion Coefficient of n-Hexane from 223 to 333 K and up to 400 MPa. J. Chern. Soc., Faraday Trans. 1, 1982, 78, 2265-2274. https://doi.org/10.1039/F19827802265
51. Mopsik FI. Dielectric Constant of N-Hexane as a Function of Temperature, Pressure, and Density. J Res Natl Bur Stand A Phys Chem. 71A (1967) 287-292. 10.6028/jres.071a.035
52. V. I. Arkhipov, N. Agmon, Relation between macroscopic and microscopic dielectric relaxation times in water dynamics  Israel Journal of Chemistry 43 (2003) 363 – 371. 10.1560/5WKJ-WJ9F-Q0DR-WPFH
53. J.K. Gregory, et al., The Water Dipole Moment in Water Clusters, Science. 275 (1997) 814-7.  10.1126/science.275.5301.814
54. B. Cabane, R. Vuilleumier, The physics of liquid water. Comptes Rendus Geoscience, 337 (2005) 159-171. 10.1016/j.crte.2004.09.018
55. J. Teixeira, Deciphering water's dielectric constant, Physics, 9 (2016) 122. 10.1103/Physics.9.122
56. A. Pasquarello, R. Resta, Dynamical monopoles and dipoles in a condensed molecular system: The case of liquid water. Phys. Rev. B, 68 (2003) 174302. 10.1103/PhysRevB.68.174302
57. M. D. Danford, H. A. Levy, Structure of water at room temperature, JACS. 84 (1962) 3965 https://doi.org/10.1021/ja00879a035.
58. J. Russo, F. Romano, H.Tanaka, New metastable form of ice and its role in the homogeneous crystallization of water. Nature Materials, 13 (2014) 733. 0.1038/nmat3977
59. C. Pérez, et al., Structures of cage, prism, and book isomers of water hexamer from broadband rotational spectroscopy, Science. 336 (2012) 897-901. 10.1126/science.1220574
60. A. Priyadarshini, A. Biswas, D. Chakraborty, B. S. Mallik. Structural and Thermophysical Anomalies of Liquid Water: A Tale of Molecules in the Instantaneous Low- and High-Density Regions.  J. Phys. Chem. B. 124 (2020) 1071. https://doi.org/10.1021/acs.jpcb.9b11596
61. K. Gets, et al., Transformation of hydrogen bond network during $CO_2$ clathrate hydrate dissociation. Appl. Surface Sci. 499 (2020) [143644].  https://doi.org/10.1016/j.apsusc.2019.143644
62. A. Shimkevich, I. Shimkevich, On Water Density Fluctuations with Helices of Hydrogen Bonds, Adv. Condens. Matter Phys. 5 (2011) (Article ID 871231). 10.1155/2011/871231.
63. E. Harder, J. D. Eaves, A. Tokmakoff, B. J. Berne, Polarizable molecules in the vibrational spectroscopy of water. Proc. Natl. Acad. Sci. U.S.A. 102, 11611–11616 (2005). 10.1073/PNAS.0505206102

64. M. Sharma, R. Resta, R. Car, Intermolecular dynamical charge fluctuations in water: A signature of the H-Bond Network. Phys. Rev. Lett. 95, 187401 (2005). https://doi.org/10.1103/PhysRevLett.95.187401

65. S. Ashihara, et al., Ultrafast structural dynamics of water induced by dissipation of vibrational energy. J. Phys. Chem. A, 111 (2007) 743–746. https://doi.org/10.1021/jp0676538

66. R. Dettori, et al., Energy Relaxation and Thermal Diffusion in Infrared Pump–Probe Spectroscopy of Hydrogen-Bonded Liquids. J. Phys. Chem. Let. 10 (2019) 3447-3452. https://doi.org/10.1021/acs.jpclett.9b01272

67. S. Woutersen, H. J. Bakker, Resonant intermolecular transfer of vibrational energy in liquid water. Nature, 402 (1999) 507–509. https://www.nature.com/articles/990058

68. A. M. Stingel, P. B. Petersen, Interpreting Quasi-Thermal Effects in Ultrafast Spectroscopy of Hydrogen-Bonded Systems. J. Phys. Chem. A, 122 (2018) 2670–2676. 10.1021/acs.jpca.7b12372

69. K. Mazur, M. Bonn, J. Hunger, Hydrogen Bond Dynamics in Primary Alcohols: A Femtosecond Infrared Study. J. Phys. Chem. B 119 (2015) 1558–1566. 10.1021/jp509816q

70. T. Kühne, R. Khaliullin, Electronic signature of the instantaneous asymmetry in the first coordination shell of liquid water, Nat. Commun. 4 (2013) 1450. https://doi.org/10.1038/ncomms2459

71. A.C. Lawson, Physics of the lindemann melting rule, Philosophical Magazine. 89 (2009) 1757-1770. 10.1080/14786430802577916

72. Menshikov L I, Menshikov P L, Fedichev P O, Effects of action at a distance in water, Phys. Usp. **63** (2020) 440–486. 10.3367/UFNe.2020.01.038721

73. G. W. Robinson, J. Lee, M.-P. Bassez, Cooperativity in liquid water, Chem. Phys. Lett. 137 (1987) 376. 10.1016/0009-2614(87)80903-X

74. Chumaevskii MA, Rodnikova MN Some peculiarities of liquid water structure. J. Mol. Liq. 106 (2003) 167–177. 10.1016/S0167-7322(03)00105-3

75. G. Lamoureux, A. D. MacKerell Jr., B. Roux, A simple polarizable model of water based on classical Drude oscillators. J. Chem. Phys. 119 (2003) 5185. https://doi.org/10.1063/1.1598191

76. E. Guardia, I. Skarmoutsos, M. Masia, Hydrogen Bonding and Related Properties in Liquid Water: A Car–Parrinello Molecular Dynamics Simulation Study. J. Phys. Chem. B 119 (2015) 8926–8938. https://doi.org/10.1021/jp507196q

77. J.D. Bernal, A geometrical approach to the structure of liquids. Nature. 183 (1959) 141

78. J. O. Richardson, et al., Concerted hydrogen-bond breaking by quantum tunneling in the water hexamer prism, Science. 351 (2016) 1310. 10.1126/science.aae0012

79. D. Laage, J. T. Hynes, On the molecular mechanism of water reorientation, J. Phys. Chem. B. 112 (2008) 14230–14242. https://doi.org/10.1021/jp805217u

80. E. Duboué-Dijon, D. Laage, Characterization of the Local Structure in Liquid Water by Various Order Parameters, J. Phys. Chem. B. 119 (2015) 8406, https://doi.org/10.1021/acs.jpcb.5b02936

81. U. Kaatze, Water, the special liquid, J. Mol. Liq, 259 (2018) 304, 10.1016/j.molliq.2018.03.038

82. A.S. Nair, P. Banerjee, S. Sarkar, B.Bagchi, Dynamics of linear molecules in water: Translation-rotation coupling in jump motion driven diffusion, J. Chem. Phys. 151, (2019) 034301; https://doi.org/10.1063/1.5100930